\documentclass[aps,prl,preprint,tightenlines,superscriptaddress,showpacs,byrevtex,tightenlines]{revtex4}
\usepackage{graphicx} 
\usepackage{dcolumn}  
\usepackage{amsmath}  

\graphicspath{{ps/}}

\begin{document}


\def\dE{{\Delta E}}
\def\mb{{M_{\rm bc}}}
\def\Dt{\Delta t}
\def\Dz{\Delta z}
\newcommand{\dmd}{\Delta m_d}
\def\fol{f_{\rm ol}}
\def\fsig{f_{\rm sig}}
 
\newcommand{\fCP}{f_{\rm sig}}
\newcommand{\zCP}{z_{\rm sig}}
\newcommand{\tCP}{t_{\rm sig}}
\newcommand{\ftag}{f_{\rm tag}}
\newcommand{\ttag}{t_{\rm tag}}
\newcommand{\ztag}{z_{\rm tag}}
\newcommand{\cala}{{\mathcal A}}
\newcommand{\cals}{{\mathcal S}}
\def\rsigbkg{{\mathcal R}_{\rm s/b}}
\def\rsigbkgBH{{\mathcal R}_{\rm s/b}^{\rm BH}}
\def\Rsig{R_{\rm sig}}
\def\Rbkg{R_{\rm bkg}}
\def\calf{{\mathcal F}}
\def\taubz{{\tau_\bz}}
\def\taubp{{\tau_\bp}}
\def\tauc{{\tau_{\ks\pip\gamma}}}
\def\taun{{\tau_{\ks\piz\gamma}}}
\newcommand*{\fq}{\ensuremath{q}}

\def\bz{{B^0}}
\def\bzb{{\overline{B}{}^0}}
\def\bbar{{\overline{B}}}
\def\bp{{B^+}}
\def\bm{{B^-}}
\def\kz{{K^0}}
\def\ks{{K_S^0}}
\def\kp{{K^+}}
\def\km{{K^-}}
\def\pip{{\pi^+}}
\def\pim{{\pi^-}}
\def\piz{{\pi^0}}
\def\kstar{{K^\ast}}
\def\kstarz{{K^{\ast 0}}}
\def\kstarp{{K^{\ast +}}}
\def\kstarm{{K^{\ast -}}}
\def\ktwostar{{K_2^\ast}}
\def\ktwostarz{{K_2^{\ast 0}}}
\def\kstarpm{{K^{\ast\pm}}}
\def\jpsi{{J/\psi}}
\def\qq{q\bar{q}}

\def\GeV{\,{\rm GeV}}
\def\GeVc{\,{\rm GeV}/c}
\def\GeVcc{\,{\rm GeV}/c^2}
\def\MeVcc{\,{\rm MeV}/c^2}

\newcommand{\BaBar}{{\sc B\hspace*{-0.2ex}a\hspace*{-0.2ex}B\hspace*{-0.2ex}a\hspace*{-0.2ex}r}}

\def\NBBsvdI{152\times 10^6}
\def\NBB04{275\times 10^6}
\def\NBB05{386\times 10^6}
\def\NBBadd{111\times 10^6}
\def\NBBsvdII{234\times 10^6}

\newcommand{\asyme}[3]{{#1^{+#2}_{-#3}}}
\newcommand{\syme}[2]{{{#1}\pm {#2}}}
\newcommand{\asymasyme}[5]{{#1^{+#2}_{-#3}{}^{+#4}_{-#5}}}
\newcommand{\symasyme}[4]{{{#1}\pm {#2}{}^{+#3}_{-#4}}}
\newcommand{\asymsyme}[4]{{{#1}^{+#2}_{-#3}\pm {#4}}}
\newcommand{\symsyme}[3]{{{#1}\pm {#2}\pm {#3}}}
\newcommand{\asymasymeSS}[5]{{#1^{+#2}_{-#3}\mbox{(stat)}{}^{+#4}_{-#5}\mbox{(syst)}}}
\newcommand{\symasymeSS}[4]{{{#1}\pm {#2}\mbox{(stat)}{}^{+#3}_{-#4}\mbox{(syst)}}}
\newcommand{\asymsymeSS}[4]{{{#1}^{+#2}_{-#3}\mbox{(stat)}\pm {#4}\mbox{(syst)}}}
\newcommand{\symsymeSS}[3]{{{#1}\pm {#2}\mbox{(stat)}\pm {#3}\mbox{(syst)}}}

\def\NevtMRI{260}
\def\NevtVMRI{116}
\def\NevtMRII{236}
\def\NevtVMRII{120}

\def\NsigMRI{\syme{70}{11}}
\def\NqqMRI{32}
\def\NrareMRI{\syme{9}{2}}
\def\PurityMRI{\syme{63}{11}}

\def\NsigMRII{\syme{45}{11}}
\def\NqqMRII{53}
\def\NrareMRII{\syme{11}{4}}
\def\PurityMRII{\syme{41}{11}}

\def\controllife{\asyme{1.524}{0.127}{0.119}}

\def\controlSin{\asyme{0.238}{0.220}{0.223}}
\def\controlCos{\syme{0.012}{0.135}}

\def\signallife{\asyme{1.474}{0.231}{0.204}}

\def\SkspizgmVal{+0.08}
\def\SkspizgmStat{0.41}
\def\SkspizgmSyst{0.10}
\def\AkspizgmVal{+0.12}
\def\AkspizgmStat{0.27}
\def\AkspizgmSyst{0.10}
\def\SkspizgmResult{\symsyme{\SkspizgmVal}{\SkspizgmStat}{\SkspizgmSyst}}
\def\SkspizgmResultSS{\symsymeSS{\SkspizgmVal}{\SkspizgmStat}{\SkspizgmSyst}}
\def\AkspizgmResult{\symsyme{\AkspizgmVal}{\AkspizgmStat}{\AkspizgmSyst}}
\def\AkspizgmResultSS{\symsymeSS{\AkspizgmVal}{\AkspizgmStat}{\AkspizgmSyst}}

\def\SkspizgmMRIVal{+0.01}
\def\SkspizgmMRIStat{0.52}
\def\SkspizgmMRISyst{0.11}
\def\AkspizgmMRIVal{+0.11}
\def\AkspizgmMRIStat{0.33}
\def\AkspizgmMRISyst{0.09}

\def\SkspizgmMRIIVal{+0.20}
\def\SkspizgmMRIIStat{0.66}
\def\AkspizgmMRIIVal{+0.14}
\def\AkspizgmMRIIStat{0.46}


\def\SkstarzgmVallast{-0.79}
\def\SkstarzgmStatplast{0.63}
\def\SkstarzgmStatnlast{0.50}
\def\SkstarzgmSystlast{0.10}
\def\SkstarzgmResultlast{\asymsyme{\SkstarzgmVallast}{\SkstarzgmStatplast}{\SkstarzgmStatnlast}{\SkstarzgmSystlast}}

\def\SkspizgmVallast{-0.58}
\def\SkspizgmStatplast{0.46}
\def\SkspizgmStatnlast{0.38}
\def\SkspizgmSystlast{0.11}
\def\SkspizgmResultlast{\asymsyme{\SkspizgmVallast}{\SkspizgmStatplast}{\SkspizgmStatnlast}{\SkspizgmSystlast}}
\def\AkspizgmVallast{+0.03}
\def\AkspizgmStatlast{0.34}
\def\AkspizgmSystlast{0.11}
\def\AkspizgmResultlast{\symsyme{\AkspizgmVallast}{\AkspizgmStatlast}{\AkspizgmSystlast}}


\preprint{\vbox{ \hbox{   }
                 \hbox{BELLE-CONF-0570}
                 \hbox{LP2005-205}
                 \hbox{EPS05-545}
}}

\title{ \quad\\[0.5cm] Time-Dependent 
{\boldmath $CP$} Asymmetries in $\bz\to\ks\piz\gamma$ transition\\
}
\date{\today}

\affiliation{Aomori University, Aomori}
\affiliation{Budker Institute of Nuclear Physics, Novosibirsk}
\affiliation{Chiba University, Chiba}
\affiliation{Chonnam National University, Kwangju}
\affiliation{University of Cincinnati, Cincinnati, Ohio 45221}
\affiliation{University of Frankfurt, Frankfurt}
\affiliation{Gyeongsang National University, Chinju}
\affiliation{University of Hawaii, Honolulu, Hawaii 96822}
\affiliation{High Energy Accelerator Research Organization (KEK), Tsukuba}
\affiliation{Hiroshima Institute of Technology, Hiroshima}
\affiliation{Institute of High Energy Physics, Chinese Academy of Sciences, Beijing}
\affiliation{Institute of High Energy Physics, Vienna}
\affiliation{Institute for Theoretical and Experimental Physics, Moscow}
\affiliation{J. Stefan Institute, Ljubljana}
\affiliation{Kanagawa University, Yokohama}
\affiliation{Korea University, Seoul}
\affiliation{Kyoto University, Kyoto}
\affiliation{Kyungpook National University, Taegu}
\affiliation{Swiss Federal Institute of Technology of Lausanne, EPFL, Lausanne}
\affiliation{University of Ljubljana, Ljubljana}
\affiliation{University of Maribor, Maribor}
\affiliation{University of Melbourne, Victoria}
\affiliation{Nagoya University, Nagoya}
\affiliation{Nara Women's University, Nara}
\affiliation{National Central University, Chung-li}
\affiliation{National Kaohsiung Normal University, Kaohsiung}
\affiliation{National United University, Miao Li}
\affiliation{Department of Physics, National Taiwan University, Taipei}
\affiliation{H. Niewodniczanski Institute of Nuclear Physics, Krakow}
\affiliation{Nippon Dental University, Niigata}
\affiliation{Niigata University, Niigata}
\affiliation{Nova Gorica Polytechnic, Nova Gorica}
\affiliation{Osaka City University, Osaka}
\affiliation{Osaka University, Osaka}
\affiliation{Panjab University, Chandigarh}
\affiliation{Peking University, Beijing}
\affiliation{Princeton University, Princeton, New Jersey 08544}
\affiliation{RIKEN BNL Research Center, Upton, New York 11973}
\affiliation{Saga University, Saga}
\affiliation{University of Science and Technology of China, Hefei}
\affiliation{Seoul National University, Seoul}
\affiliation{Shinshu University, Nagano}
\affiliation{Sungkyunkwan University, Suwon}
\affiliation{University of Sydney, Sydney NSW}
\affiliation{Tata Institute of Fundamental Research, Bombay}
\affiliation{Toho University, Funabashi}
\affiliation{Tohoku Gakuin University, Tagajo}
\affiliation{Tohoku University, Sendai}
\affiliation{Department of Physics, University of Tokyo, Tokyo}
\affiliation{Tokyo Institute of Technology, Tokyo}
\affiliation{Tokyo Metropolitan University, Tokyo}
\affiliation{Tokyo University of Agriculture and Technology, Tokyo}
\affiliation{Toyama National College of Maritime Technology, Toyama}
\affiliation{University of Tsukuba, Tsukuba}
\affiliation{Utkal University, Bhubaneswer}
\affiliation{Virginia Polytechnic Institute and State University, Blacksburg, Virginia 24061}
\affiliation{Yonsei University, Seoul}
  \author{K.~Abe}\affiliation{High Energy Accelerator Research Organization (KEK), Tsukuba} 
  \author{K.~Abe}\affiliation{Tohoku Gakuin University, Tagajo} 
  \author{I.~Adachi}\affiliation{High Energy Accelerator Research Organization (KEK), Tsukuba} 
  \author{H.~Aihara}\affiliation{Department of Physics, University of Tokyo, Tokyo} 
  \author{K.~Aoki}\affiliation{Nagoya University, Nagoya} 
  \author{K.~Arinstein}\affiliation{Budker Institute of Nuclear Physics, Novosibirsk} 
  \author{Y.~Asano}\affiliation{University of Tsukuba, Tsukuba} 
  \author{T.~Aso}\affiliation{Toyama National College of Maritime Technology, Toyama} 
  \author{V.~Aulchenko}\affiliation{Budker Institute of Nuclear Physics, Novosibirsk} 
  \author{T.~Aushev}\affiliation{Institute for Theoretical and Experimental Physics, Moscow} 
% ...
\collaboration{The Belle Collaboration}

\collaboration{Belle Collaboration}
\noaffiliation

\begin{abstract}
 We present measurements of $CP$-violation parameters in $b \to s\gamma$
 transitions based on a sample of $\NBB05$ $B\bbar$ pairs collected at
 the $\Upsilon(4S)$ resonance with the Belle detector at the KEKB
 energy-asymmetric $e^+e^-$ collider.  One neutral $B$ meson is fully
 reconstructed in the $\bz\to\ks\piz\gamma$ decay channel irrespective to
 the $\ks\piz$ intermediate state.
  The flavor of the accompanying $B$ meson is identified from its decay
 products. $CP$-violation parameters are obtained from the
 asymmetries in the distributions of the proper-time intervals between
 the two $B$ decays.

 We obtain the following results
 for the $\ks\piz$ invariant mass covering the full range up to
 $1.8\GeVcc$:
\begin{eqnarray}
 \cals_{\ks\piz\gamma}&=&\SkspizgmResultSS,\nonumber\\
 \cala_{\ks\piz\gamma}&=&\AkspizgmResultSS.\nonumber
\end{eqnarray}

\end{abstract}


\maketitle


{\renewcommand{\thefootnote}{\fnsymbol{footnote}}}
\setcounter{footnote}{0}

\section{Introduction}

In the Standard Model (SM), $CP$ violation arises from an irreducible
phase, the Kobayashi-Maskawa (KM) phase~\cite{Kobayashi:1973fv}, in the
weak-interaction quark-mixing matrix.
The phenomena of time-dependent $CP$ violation in decays through
radiative penguin processes such as $b\to s\gamma$ are sensitive to
physics beyond the SM.
Within the SM, the photon emitted from a $\bz$ ($\bzb$)
meson is dominantly right-handed (left-handed).
Therefore the polarization of the photon carries
information on the original $b$-flavor and the decay
is, thus, almost flavor-specific.
As a result, the SM predicts 
a small asymmetry~\cite{Atwood:1997zr,Grinstein:2004uu} and
any significant deviation from this expectation
would be a manifestation of new physics.
It was pointed out that 
in decays of the type $\bz\to P^0Q^0\gamma$, where $P^0$ and $Q^0$
represent any $CP$ eigenstate spin-0 neutral particles
(e.g. $P^0 = \ks$ and $Q^0 = \piz$),
many new physics effects on the mixing-induced 
$CP$ violation do not depend on the resonant structure
of the $P^0Q^0$ system~\cite{Atwood:2004jj}.

At the KEKB energy-asymmetric $e^+e^-$ (3.5 on 8.0$\GeV$)
collider~\cite{bib:KEKB}, the $\Upsilon(4S)$ is produced with a Lorentz
boost of $\beta\gamma=0.425$ along the $z$ axis, which is defined as the
direction antiparallel
to the $e^+$ beam direction.
In the decay chain $\Upsilon(4S)\to \bz\bzb \to \fCP \ftag$, where one
of the $B$ mesons decays at time $\tCP$ to a final state $\fCP$, which
is our signal mode, and the other decays at time $\ttag$ to a final
state $\ftag$ that distinguishes between $B^0$ and $\bzb$, the decay
rate has a time dependence given by
\begin{eqnarray}
\label{eq:psig}
{\cal P}(\Dt) = 
\frac{e^{-|\Dt|/{\taubz}}}{4{\taubz}}
\biggl\{1 + \fq
\Bigl[ \cals\sin(\dmd\Dt)
   + \cala\cos(\dmd\Dt)
\Bigr]
\biggr\}.
\end{eqnarray}
Here $\cals$ and $\cala$ are $CP$-violation parameters, $\taubz$ is the
$B^0$ lifetime, $\dmd$ is the mass difference between the two $B^0$ mass
eigenstates, $\Dt$ is the time difference $\tCP - \ttag$, and the
$b$-flavor charge $\fq$ = +1 ($-1$) when the tagging $B$ meson is a
$B^0$ ($\bzb$).
Since the $B^0$ and $\bzb$ mesons are approximately at 
rest in the $\Upsilon(4S)$ center-of-mass system (c.m.s.),
$\Dt$ can be determined from the displacement in $z$ 
between the $\fCP$ and $f_{\rm tag}$ decay vertices:
$\Delta t \simeq (\zCP - \ztag)/(\beta\gamma c)
 \equiv \Delta z/(\beta\gamma c)$.

For $\bz\to\ks\piz\gamma$, the $\ks$ vertex is displaced from the $B$
vertex and often lies outside of the silicon vertex detector (SVD).
When the $\ks$ vertex can be reconstructed inside the SVD, the
time-dependent $CP$ asymmetry can be measured.
Measurements of such $CP$ asymmetries were previously 
reported by \BaBar~\cite{Aubert:2004pe} and
Belle~\cite{bib:ushiroda05} in the $\bz\to\kstarz(\to\ks\piz)\gamma$ decay:
\begin{eqnarray*}
 \cals_{\kstarz\gamma} &=& 0.25 \pm 0.63 \pm 0.14 \phantom{{\mbox {\rm Belle}}}
({\mbox \BaBar})\\
 \cals_{\kstarz\gamma} &=& \SkstarzgmResultlast \phantom{{\mbox \BaBar}}
({\mbox {\rm Belle}}).
\end{eqnarray*}
Belle also measured these asymmetries with an extended $M_{\ks\piz}$ mass
region\cite{bib:ushiroda05} ($M_{\ks\piz}<1.8\GeVcc$):
\[
  \cals_{\ks\piz\gamma} = \SkspizgmResultlast \phantom{{\mbox \BaBar}}
({\mbox {\rm Belle}}).
\]
In this analysis, we update the $CP$ measurements for
$\bz\to\ks\piz\gamma$ in the mass region $M_{\ks\piz}<1.8\GeVcc$
with an additional dataset of $\NBBadd$ $B\bbar$ pairs.

The Belle detector is a large-solid-angle magnetic
spectrometer that
consists of an SVD,
a 50-layer central drift chamber (CDC), an array of
aerogel threshold \v{C}erenkov counters (ACC), 
a barrel-like arrangement of time-of-flight
scintillation counters (TOF), and an electromagnetic calorimeter
comprised of CsI(Tl) crystals (ECL) located inside 
a super-conducting solenoid coil that provides a 1.5~T
magnetic field.  An iron flux-return located outside of
the coil is instrumented to detect $K_L^0$ mesons and to identify
muons (KLM).  The detector
is described in detail elsewhere~\cite{Belle}.
Two inner detector configurations were used. A 2.0 cm beampipe
and a 3-layer silicon vertex detector (SVD1) was used for the first sample
of $\NBBsvdI$ $B\bbar$ pairs, while a 1.5 cm beampipe, a 4-layer
silicon detector (SVD2) and a small-cell inner drift chamber were used to record  
the remaining $\NBBsvdII$ $B\bbar$ pairs~\cite{Ushiroda}.

\section{Event Selection, Flavor Tagging and Vertex Reconstruction}
\subsection{Event Selection for $\ks\piz\gamma$}

For high energy prompt photons, we select an isolated cluster in the ECL
that has no corresponding charged track, and has the largest energy in
the c.m.s. We require the shower shape to be consistent with that of a
photon.  In order to reduce the background from $\piz$ and $\eta$ mesons, we
exclude photons compatible with $\piz\to\gamma\gamma$ or
$\eta\to\gamma\gamma$ decays; we reject photon pairs that satisfy
$\mathcal{L}_{\piz}\ge 0.18$ or $\mathcal{L}_{\eta}\ge 0.18$, where
$\mathcal{L}_{\piz(\eta)}$ is a $\piz$ ($\eta$) likelihood described in
detail elsewhere~\cite{Koppenburg:2004fz}.  The polar angle of the
photon direction in the laboratory frame is restricted to the barrel
region of the ECL ($33^\circ < \theta_\gamma < 128^\circ$), but is
extended to the end-cap regions ($17^\circ < \theta_\gamma < 150^\circ$)
for the second data sample due to the reduced material in front of the
ECL.

Neutral kaons ($\ks$) are reconstructed from two oppositely charged
pions that have an invariant mass within $\pm 6\MeVcc$ ($2\sigma$) of
the $\ks$ nominal mass.  The $\pip\pim$ vertex is required to be
displaced from the interaction point (IP) in the direction of the pion
pair momentum~\cite{Abe:2004xp}.  Neutral pions ($\piz$) are formed from
two photons with the invariant mass within $\pm 16\MeVcc$ ($3\sigma$) of
the $\piz$ mass.  The photon momenta are then recalculated with a $\piz$
mass constraint and we require the momentum of $\piz$ candidates in the
c.m.s. to be greater than $0.3\GeVc$.  The $\ks\piz$ invariant
mass, $M_{\ks\piz}$, is required to be less than $1.8\GeVcc$.

$\bz$ mesons are reconstructed by combining $\ks$, $\piz$ and $\gamma$
candidates. We form two kinematic variables: the energy difference
$\dE\equiv E_B^{\rm c.m.s.}-E_{\rm beam}^{\rm c.m.s.}$ and the beam-energy
constrained mass $\mb\equiv\sqrt{(E_{\rm beam}^{\rm c.m.s.})^2-(p_B^{\rm
c.m.s.})^2}$, where $E_{\rm beam}^{\rm c.m.s.}$ is the beam energy, $E_B^{\rm
c.m.s.}$ and $p_B^{\rm c.m.s.}$ are the energy and the momentum of the
candidate in the c.m.s.  Candidates are accepted if they have $\mb >
5.2\GeVcc$ and $-0.5\GeV < \dE < 0.5\GeV$.

We reconstruct $\bp\to\ks\pip\gamma$ candidates in a similar way as the
$\bz\to\ks\piz\gamma$ decay in order to reduce the cross-feed background
from $\bp\to\ks\pip\gamma$ in $\bz\to\ks\piz\gamma$.  The
$\bp\to\ks\pip\gamma$ events are also used for various crosschecks.
For a $\pip$ candidate, we require that the track originates from the IP
and that the transverse momentum is greater than $0.1\GeVc$.  We also
require that the $\pip$ candidate cannot be identified as any other
particle species ($K^+, p^+, e^+,$ and $\mu^+$).

Candidate $\bp\to\ks\pip\gamma$ and $\bz\to\ks\piz\gamma$ decays are
selected simultaneously; we allow only one candidate for each event.
The best candidate selection is based on the event likelihood ratio
$\rsigbkg$ that is obtained from a Fisher discriminant
$\calf$~\cite{Fisher}, which uses the extended modified Fox-Wolfram
moments~\cite{Abe:2003yy} as discriminating variables.
We select the candidate that has the largest $\rsigbkg$.
The signal region is defined as $-0.2\GeV < \dE < 0.1\GeV$ and
$5.27\GeVcc < \mb < 5.29\GeVcc$.

We use events outside the signal region as well as large Monte Carlo
(MC) samples to study the background components.  The dominant
background is from continuum light quark pair production ($e^+e^-\to
q\,\bar{q}$ with $q = u,d,s,c$), which we refer to as $\qq$ hereafter.
In order to reduce the $\qq$ background contribution, we form another
event likelihood ratio $\rsigbkgBH$ by combining $\rsigbkg$ with
$\cos\theta_H$ and $\cos\theta_B$, where $\theta_B$ is the polar angle
of the $B$ meson candidate momentum in the laboratory frame, and $\theta_H$
is the angle between the $B$ candidate momentum and the daughter
$\ks$ momentum in the rest frame of the $\ks\pi$ system.
Since the relative background contribution will be smaller in the region
of the $K^*$, we introduce two $\ks\piz$ invariant mass regions: MR1,
defined as $0.8\GeVcc < M_{\ks\piz} < 1.0\GeVcc$, and MR2 which is
defined as $M_{\ks\piz} < 1.8\GeVcc$ after excluding MR1. The specific
$\rsigbkgBH$ selection criteria applied depend on both the mass region
and flavor tagging information.
After applying all other selection criteria described so far,
77\% of the $\qq$ background is rejected while 87\% of
the $\kstarz\gamma$ signal is retained in MR1; in MR2, 87\% of the $\qq$
is rejected while 68\% of the $\ktwostarz\gamma$ signal is retained.
Background contributions
from $B$ decays, which are considerably smaller than $\qq$, are
dominated by cross-feed from other radiative $B$ decays including
$\bp\to\ks\pip\gamma$.

\subsection{Flavor Tagging}

The $b$-flavor of the accompanying $B$ meson is identified from
inclusive properties of particles that are not associated with the
reconstructed signal decay.  The algorithm for flavor tagging is
described in detail elsewhere~\cite{bib:fbtg_nim}.  We use two
parameters, $\fq$ defined in Eq.~(\ref{eq:psig}) and $r$, to represent
the tagging information.  The parameter $r$ is an event-by-event
flavor-tagging dilution factor that ranges from 0 to 1; $r=0$ when there
is no flavor discrimination and $r=1$ implies unambiguous flavor
assignment.  It is determined by using MC data and is only used to sort
data into six $r$ intervals.  The wrong tag fraction $w$ and the
difference $\Delta w$ in $w$ between the $\bz$ and $\bzb$ decays are
determined for each of the six $r$ intervals from
data~\cite{Abe:2004xp}.

\subsection{Vertex Reconstruction}

The vertex position of the signal-side decay is reconstructed from the
$\ks$ trajectory with a constraint on the IP; the IP profile
($\sigma_x\simeq 100\rm\,\mu m$, $\sigma_y\simeq 5\rm\,\mu m$,
$\sigma_z\simeq 3\rm\,mm$) is convolved with the finite $B$ flight
length in the plane perpendicular to the $z$ axis.  Both pions from the
$\ks$ decay are required to have enough hits in the SVD in order to
reconstruct the $\ks$ trajectory with high resolution:
at least one layer with hits on both sides and at least one additional
hit in the $z$ side of the other layers for SVD1, and at least two layers
with hits on both sides for SVD2.
  The reconstruction efficiency depends not only on the $\ks$ momentum
  but also on the SVD geometry.  The efficiency with SVD2 (51\%) is
significantly higher
than with SVD1 (40\%) because of the larger detection volume.  The other
(tag-side) $B$ vertex determination is the same as that for the
$\bz\to\phi\ks$ analysis~\cite{Abe:2004xp}.

\section{Signal Yield Extraction}

\begin{figure}
\resizebox{0.46\columnwidth}{!}{\includegraphics{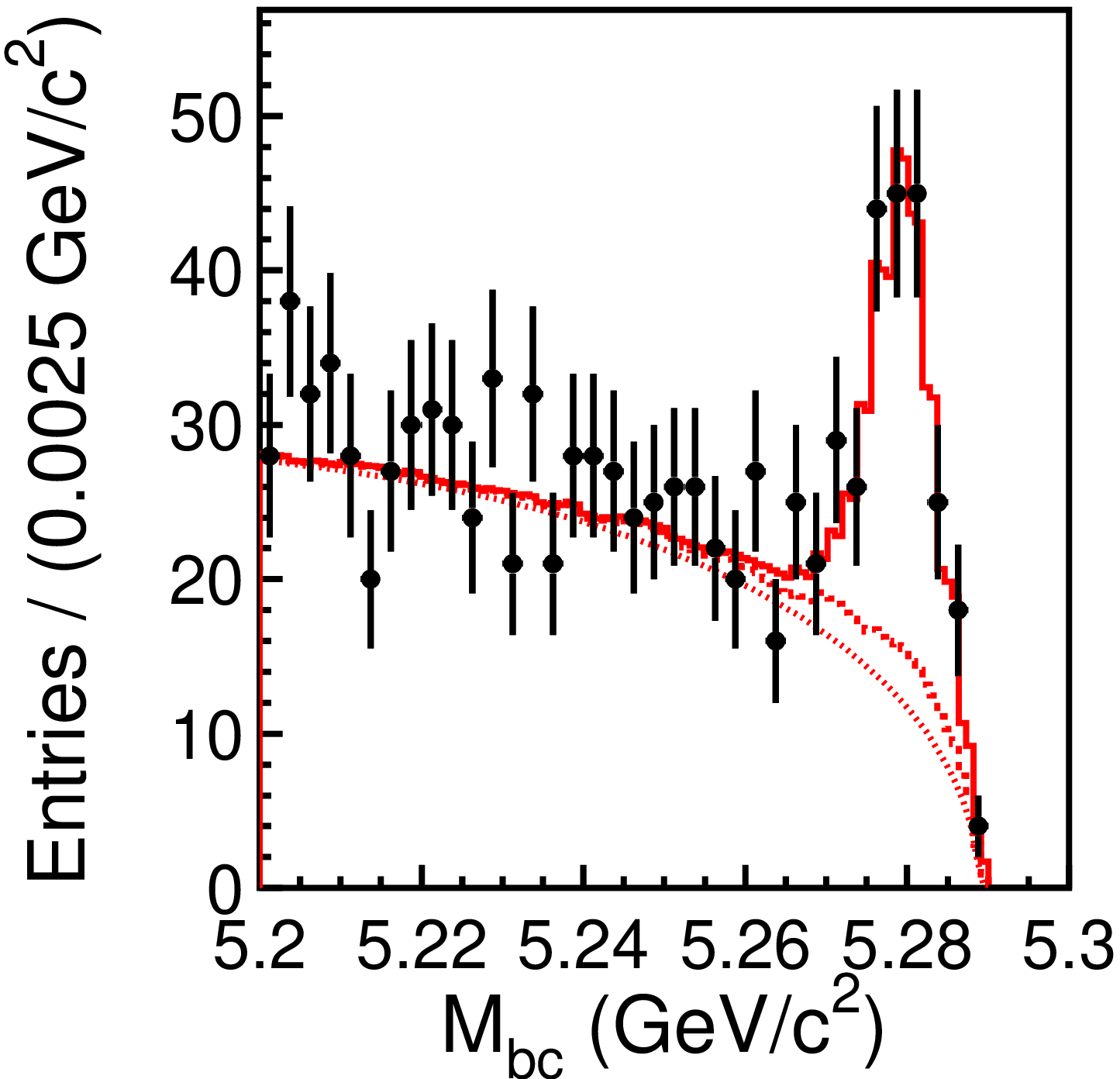}}
\resizebox{0.46\columnwidth}{!}{\includegraphics{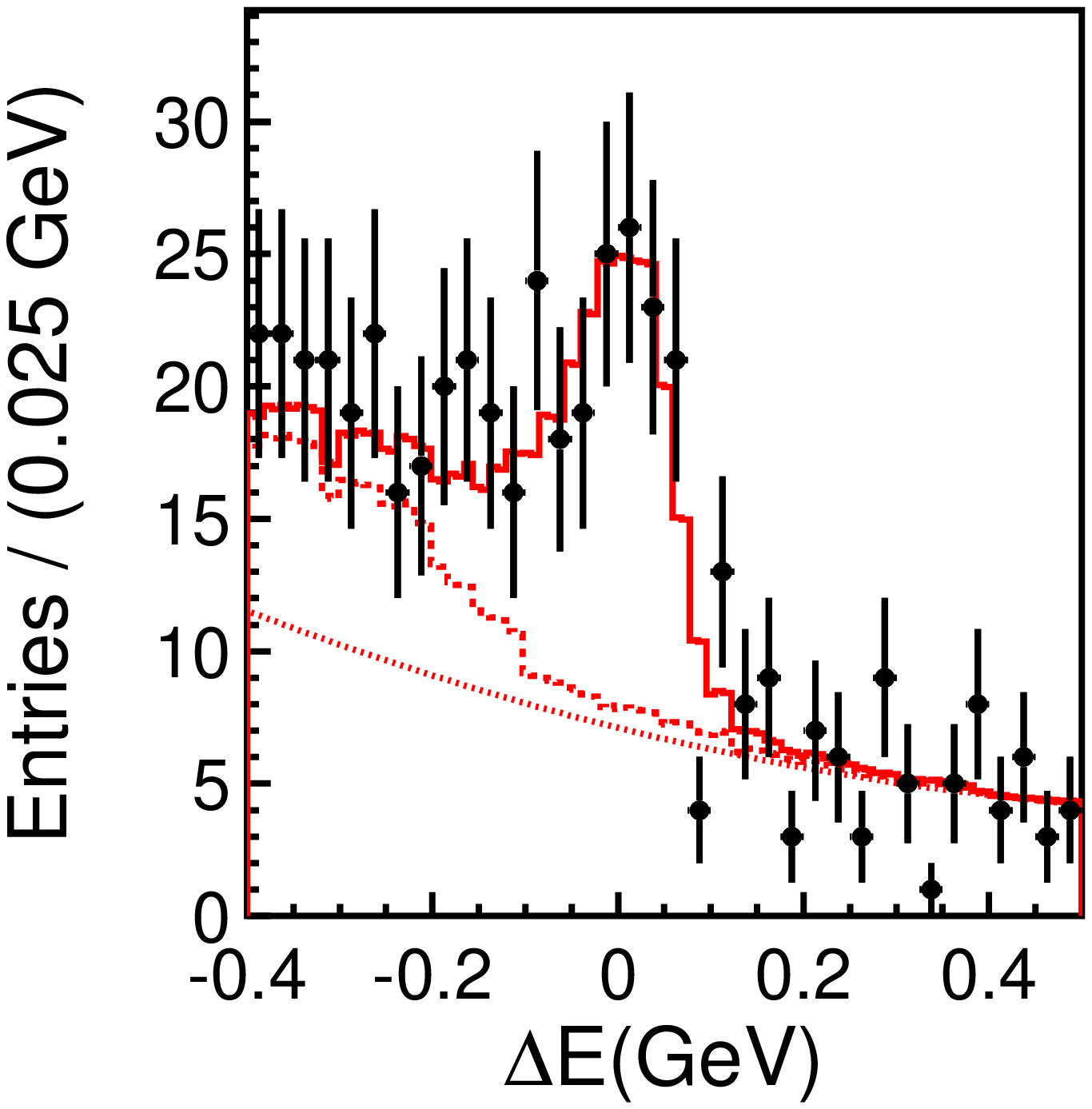}}
\caption{(a) $\mb$ distributions within the $\dE$ signal region
 and (b) $\dE$ distributions within the $\mb$ signal (MR1 and MR2 combined).
 Solid curves show the fit to signal plus background distributions.
 Lower dashed curves show the background contributions from $\qq$.
 Upper dashed histogram show the sum of background contributions from
 $\qq$ and $B$ decays.
 }
\label{fig:mb}
\end{figure}

Figure~\ref{fig:mb} shows the $\mb$ ($\dE$) distribution for the
reconstructed $\ks\piz\gamma$ candidates within the $\dE$ ($\mb$) signal
region after flavor tagging and vertex reconstruction.  The signal yield
is determined from an unbinned two-dimensional maximum-likelihood fit to
the $\dE$-$\mb$ distribution.  The fit region is chosen as $-0.4\GeV <
\dE < 0.5\GeV$ and $5.2\GeVcc < \mb$ to avoid other $B\bbar$ background
events that populate the low-$\dE$ high-$M_{\ks\piz}$ region. The
signal distribution is represented by
a PDF obtained from an MC simulation of
$\bz\to\kstarz\gamma$ and $\bz\to\ktwostarz\gamma$ that accounts for a
small correlation between $\mb$ and $\dE$.  The background from $B$
decays are also modeled with an MC simulation.  For the $\qq$ background,
we use the ARGUS parameterization~\cite{bib:ARGUS} for $\mb$ and a
second-order polynomial for $\dE$.  The normalizations of the signal and
background distributions and the $\qq$ background shape are the five
free parameters in the fit.  We observe a total of $\NevtMRI$ candidates in
the signal box in MR1, which decreases to $\NevtVMRI$ after flavor
tagging and $B$ vertex reconstruction, and obtain $\NsigMRI$ signal
events from the fit; the average signal purity over the six $r$ intervals
is $\PurityMRI$\%.  In MR2, corresponding numbers are $\NevtMRII$,
$\NevtVMRII$, $\NsigMRII$, and $\PurityMRII$\%.


\section{{\boldmath $CP$} Asymmetry Measurements}

We determine $\cals$ and $\cala$ from an unbinned maximum-likelihood fit
to the observed $\Dt$ distribution.  The probability density function
(PDF) expected for the signal distribution, ${\cal P}_{\rm
sig}(\Dt;\cals,\cala,\fq,w,\Delta w)$, is given by the time dependent
decay rate [Eq.~(\ref{eq:psig})] modified to incorporate the effect of
incorrect flavor assignment. The distribution is convolved with the
proper-time interval resolution function $\Rsig$, which takes into
account the finite vertex resolution.  The parametrization of $\Rsig$ is
the same as that used for the $\bz\to\ks\piz$
decay~\cite{Abe:2004xp}. $\Rsig$ is first derived from flavor-specific
$B$ decays~\cite{bib:BELLE-CONF-0436} and modified by additional
parameters that rescale vertex errors to account for the fact that
there is no track directly originating from the $B$ meson decay point.

For each event, the following likelihood function is
evaluated:
\begin{equation}
  \begin{split}
      P_i
      =& (1-\fol)\int_{-\infty}^{+\infty} \biggl[
      \fsig{\cal P}_{\rm sig}(\Dt')\Rsig (\Dt_i-\Dt') \\
      &+(1-\fsig){\cal P}_{\rm bkg}(\Dt')\Rbkg (\Dt_i-\Dt')\biggr]
      d(\Dt')  \\
      &+\fol P_{\rm ol}(\Dt_i),
      \label{eq:likelihood}
  \end{split}
\end{equation}
where $P_{\rm ol}$ is a Gaussian function that represents a small
outlier component with fraction $\fol$~\cite{bib:resol}.  The signal
probability $\fsig$ is calculated on an event-by-event basis from the
function which we obtained as the result of the two-dimensional
$\dE$-$\mb$ fit for the signal yield extraction.  A PDF for background
events, ${\cal P}_{\rm bkg}$, is modeled as a sum of exponential and
prompt components, and is convolved with a Gaussian which
represents the resolution function $\Rbkg$ for the background.  All
parameters in ${\cal P}_{\rm bkg}$ and $\Rbkg$ are determined by a fit
to the $\Dt$ distribution of a background-enhanced control sample,
i.e. events outside of the $\dE$-$\mb$ signal region.  We fix $\tau_\bz$
and $\dmd$ at their world-average values~\cite{bib:HFAG}.

The only free parameters in the final fit are $\cals_{\ks\piz\gamma}$
and $\cala_{\ks\piz\gamma}$, which are
determined by maximizing the likelihood function $L =
\prod_iP_i(\Dt_i;\cals,\cala)$ where the product is over all events.
We obtain
\begin{eqnarray}
 \cals_{\ks\piz\gamma} &=& \SkspizgmResultSS,\nonumber \\
 \cala_{\ks\piz\gamma} &=& \AkspizgmResultSS.\nonumber
\end{eqnarray}

\begin{figure}
\resizebox{0.46\columnwidth}{!}{\includegraphics{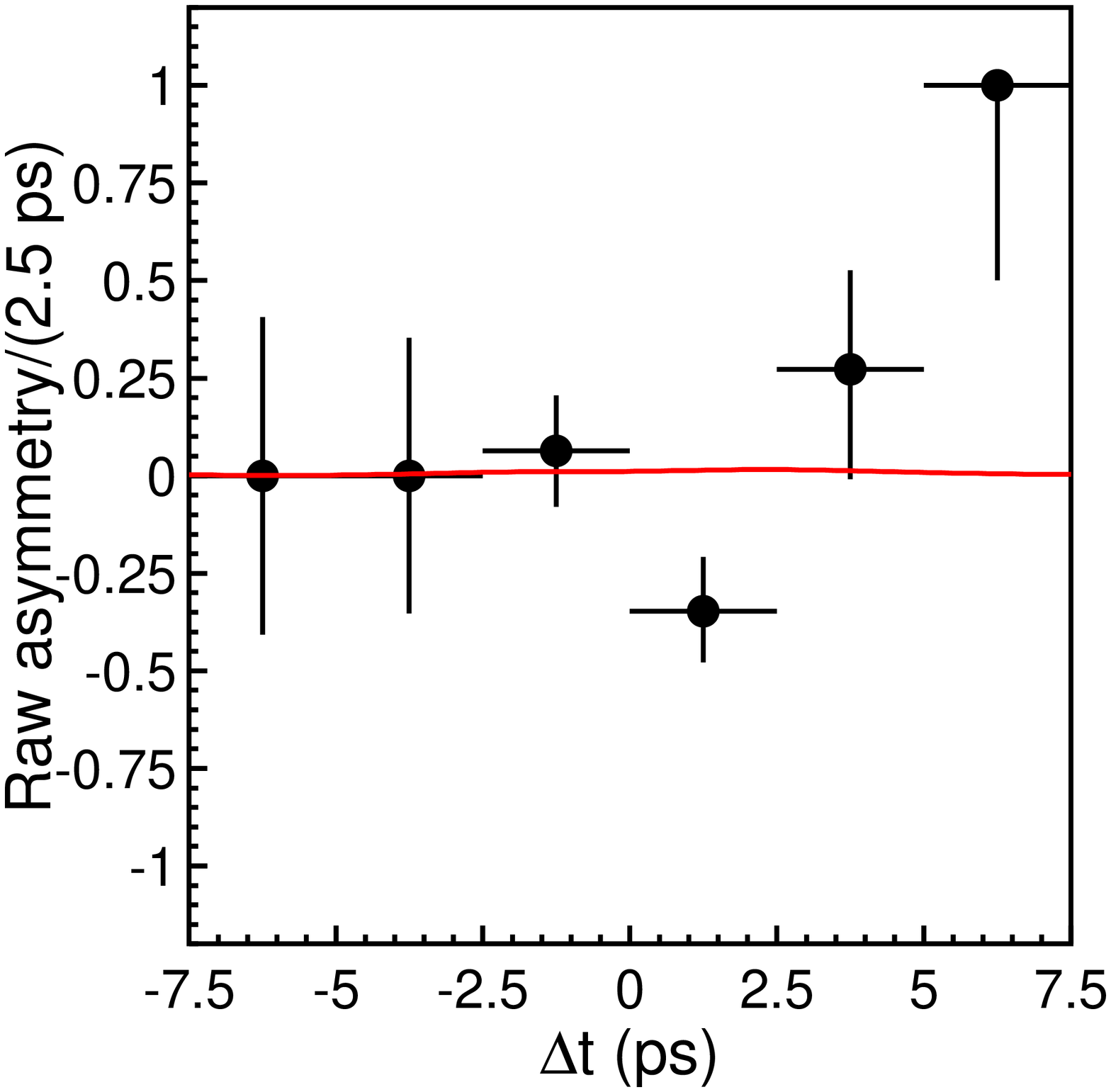}}
\resizebox{0.46\columnwidth}{!}{\includegraphics{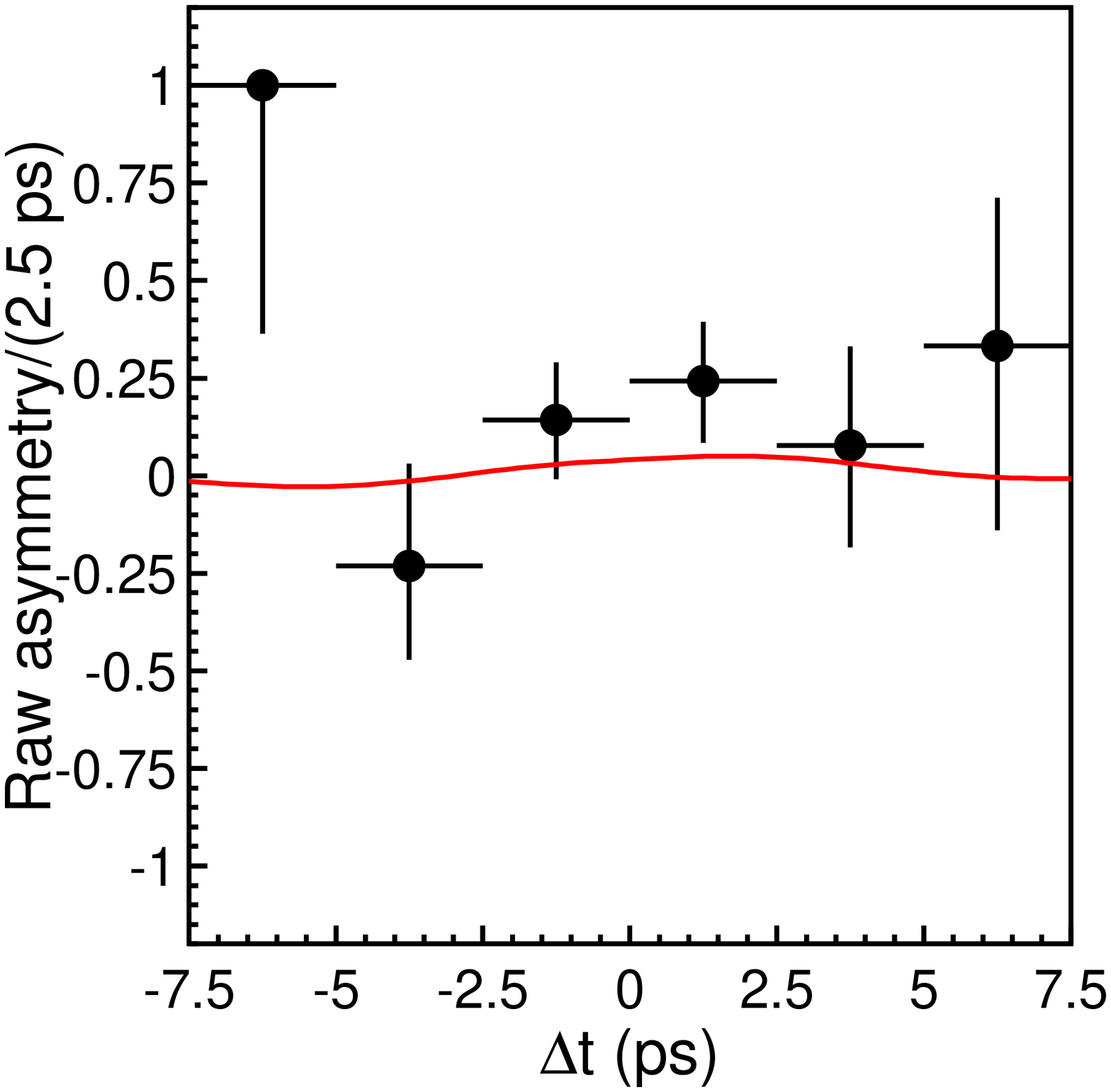}}
\caption{
 Asymmetry in each $\Dt$ bin for $\bz\to\ks\piz\gamma$ with $0 < r \le
 0.5$(left) and $0.5 < r\le 1.0$(right).\\
 Solid curves show the results of the unbinned maximum-likelihood
 fits.
 }
\label{fig:asym}
\end{figure}

We define the raw asymmetry in each $\Dt$ bin by
$(N_{q=+1}-N_{q=-1})/(N_{q=+1}+N_{q=-1})$, where $N_{q=+1(-1)}$ is the
number of observed candidates with $q=+1(-1)$.  Figure~\ref{fig:asym}
shows the raw asymmetries for the $\ks\piz\gamma$ events.  Note that
these are simple projections onto the $\Delta t$ axis, and do not
reflect other event-by-event information (such as the signal fraction,
the wrong tag fraction and the vertex resolution), which is in fact used
in the unbinned maximum-likelihood fit for $\cals$ and $\cala$.

\subsection{Systematic Error}

Primary sources of the systematic error are
(1) uncertainties in the resolution function
($\pm 0.06$ for $\cals_{\ks\piz\gamma}$ and
  $\pm 0.03$ for $\cala_{\ks\piz\gamma}$),
(2) uncertainties in the vertex reconstruction 
($\pm 0.03$ for $\cals_{\ks\piz\gamma}$ and
  $\pm 0.04$ for $\cala_{\ks\piz\gamma}$) and
(3) uncertainties in the background fraction
($\pm 0.07$ for $\cals_{\ks\piz\gamma}$ and
  $\pm 0.03$ for $\cala_{\ks\piz\gamma}$).
Effects of tag-side interference~\cite{Long:2003wq}
contribute
$\pm 0.07$ for $\cala_{\ks\piz\gamma}$.
Also included are effects 
from uncertainties in the wrong tag fraction and
physics parameters ($\dmd$, $\taubz$ and $\cala_{\kstarz\gamma}$). 
Fitting a large sample of MC events revealed no bias in the fit procedure.
The statistical errors from the MC fit are assigned as systematic
errors.
The total systematic error is obtained by adding
these contributions in quadrature.

\subsection{Crosschecks}

Various crosschecks of the measurement are performed.  We apply the
same fit procedure to the $\bz\to\jpsi\ks$ sample without using $\jpsi$
daughter tracks for the vertex reconstruction~\cite{bib:sqq05}. We obtain
$\cals_{\jpsi\ks} = +0.73\pm 0.08$(stat) and $\cala_{\jpsi\ks} =
+0.01\pm 0.04$(stat), which are in good agreement with the world-average
values~\cite{bib:PDG2004}.  We perform a fit to $\bp\to\ks\pip\gamma$
, which is a good control sample
of the $\bz\to\ks\piz\gamma$ decay, without using the primary $\pip$ for
the vertex reconstruction.
The result is consistent with no $CP$ asymmetry, as expected.  Lifetime
measurements are also performed for these modes, and values consistent
with the world-average values are obtained.  Ensemble tests are carried out
with MC pseudo-experiments using $\cals$ and $\cala$ obtained by the fit
as the input parameters. We find that the statistical errors obtained
in our measurements are all within the expectations from the ensemble
tests.  Fits to the two $M_{\ks\piz}$ regions yield
$\cals = \symsyme{\SkspizgmMRIVal}{\SkspizgmMRIStat\mbox{(stat)}}{\SkspizgmMRISyst\mbox{(syst)}}$ and
$\cala = \symsyme{\AkspizgmMRIVal}{\AkspizgmMRIStat\mbox{(stat)}}{\AkspizgmMRISyst\mbox{(syst)}}$ for MR1,
and
$\cals = \syme{\SkspizgmMRIIVal}{\SkspizgmMRIIStat}$(stat) and
$\cala = \syme{\AkspizgmMRIIVal}{\AkspizgmMRIIStat}$(stat) for MR2.
The results are consistent with those from the full $M_{\ks\piz}$ sample.

\section{Summary}

We have performed a measurement of the time-dependent $CP$ asymmetry
in the decay $\bz\to\ks\piz\gamma$ with 
$\ks\piz$ invariant mass up to $1.8\GeVcc$, based on a sample of
$\NBB05$ $B\bbar$ pairs.
We obtain $CP$-violation parameters
$\cals_{\ks\piz\gamma}=\SkspizgmResultSS$ and
$\cala_{\ks\piz\gamma}=\AkspizgmResultSS$.
We do not find any significant $CP$ asymmetry, and therefore no
indication of new physics from right handed currents, with the present
statistics.

\section{Acknowledgment}
We thank the KEKB group for the excellent operation of the
accelerator, the KEK cryogenics group for the efficient
operation of the solenoid, and the KEK computer group and
the National Institute of Informatics for valuable computing
and Super-SINET network support. We acknowledge support from
the Ministry of Education, Culture, Sports, Science, and
Technology of Japan and the Japan Society for the Promotion
of Science; the Australian Research Council and the
Australian Department of Education, Science and Training;
the National Science Foundation of China under contract
No.~10175071; the Department of Science and Technology of
India; the BK21 program of the Ministry of Education of
Korea and the CHEP SRC program of the Korea Science and
Engineering Foundation; the Polish State Committee for
Scientific Research under contract No.~2P03B 01324; the
Ministry of Science and Technology of the Russian
Federation; the Ministry of Higher Education, 
Science and Technology of the Republic of Slovenia;  
the Swiss National Science Foundation; the National Science Council and
the Ministry of Education of Taiwan; and the U.S.\
Department of Energy.

\end{document}